\documentclass[8.5pt,twoside,twocolumn]{article}
\usepackage[super,sort&compress,comma]{natbib} 
\usepackage{mhchem}
\usepackage{times,mathptmx}
\usepackage{sectsty}
\usepackage{balance} 

\usepackage{graphicx} 
\usepackage{lastpage}
\usepackage[format=plain,justification=raggedright,singlelinecheck=false,font=small,labelfont=bf,labelsep=space]{caption} 
\usepackage{fancyhdr}
\begin{document}

\thispagestyle{plain}
\fancypagestyle{plain}{
\fancyhead[L]{\includegraphics[height=8pt]{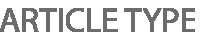}}
\fancyhead[C]{\hspace{-1cm}\includegraphics[height=20pt]{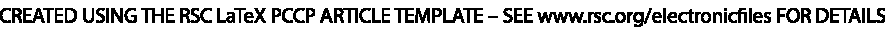}}
\fancyhead[R]{\includegraphics[height=10pt]{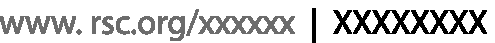}\vspace{-0.2cm}}
\renewcommand{\headrulewidth}{1pt}}
\renewcommand{\thefootnote}{\fnsymbol{footnote}}
\renewcommand\footnoterule{\vspace*{1pt}%
\hrule width 3.4in height 0.4pt \vspace*{5pt}} 
\setcounter{secnumdepth}{5}

\makeatletter 
\def\subsubsection{\@startsection{subsubsection}{3}{10pt}{-1.25ex plus -1ex minus -.1ex}{0ex plus 0ex}{\normalsize\bf}} 
\def\paragraph{\@startsection{paragraph}{4}{10pt}{-1.25ex plus -1ex minus -.1ex}{0ex plus 0ex}{\normalsize\textit}} 
\renewcommand\@biblabel[1]{#1}            
\renewcommand\@makefntext[1]%
{\noindent\makebox[0pt][r]{\@thefnmark\,}#1}
\makeatother 
\renewcommand{\figurename}{\small{Fig.}~}
\sectionfont{\large}
\subsectionfont{\normalsize} 

\fancyfoot{}
\fancyfoot[LO,RE]{\vspace{-7pt}\includegraphics[height=9pt]{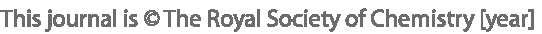}}
\fancyfoot[CO]{\vspace{-7.2pt}\hspace{12.2cm}\includegraphics{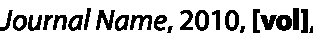}}
\fancyfoot[CE]{\vspace{-7.5pt}\hspace{-13.5cm}\includegraphics{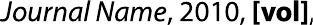}}
\fancyfoot[RO]{\footnotesize{\sffamily{1--\pageref{LastPage} ~\textbar  \hspace{2pt}\thepage}}}
\fancyfoot[LE]{\footnotesize{\sffamily{\thepage~\textbar\hspace{3.45cm} 1--\pageref{LastPage}}}}
\fancyhead{}
\renewcommand{\headrulewidth}{1pt} 
\renewcommand{\footrulewidth}{1pt}
\setlength{\arrayrulewidth}{1pt}
\setlength{\columnsep}{6.5mm}
\setlength\bibsep{1pt}

\twocolumn[
  \begin{@twocolumnfalse}
\noindent\LARGE{\textbf{Stripe Systems with Competing Interactions on Quasi-One
Dimensional Periodic Substrates }}
\vspace{0.6cm}

\noindent\large{\textbf{Danielle McDermott,\textit{$^{a,b}$} 
Cynthia J. Olson Reichhardt,$^{\ast}$\textit{$^{a}$} and
Charles Reichhardt\textit{$^{a}$}}}\vspace{0.5cm}

\noindent\textit{\small{\textbf{Received Xth XXXXXXXXXX 20XX, Accepted Xth XXXXXXXXX 20XX\newline
First published on the web Xth XXXXXXXXXX 200X}}}

\noindent \textbf{\small{DOI: 10.1039/b000000x}}
\vspace{0.6cm}

\noindent \normalsize{
We numerically examine the two-dimensional ordering of a stripe
forming system of particles
with competing long-range repulsion and short-range attraction                
in the presence of a quasi-one-dimensional corrugated substrate.
As a function of increasing substrate
strength or the ratio of the 
number of particles to the number of substrate minima
we show that a remarkable  variety of
distinct orderings can be realized, including modulated 
stripes, prolate clump phases, two dimensional ordered kink structures,
crystalline void phases, and smectic phases. 
Additionally in some cases the
stripes align perpendicular to the substrate troughs.
Our results suggest 
that a new route to self assembly for systems with competing
interactions can be achieved 
through the addition of a simple periodic modulated substrate.    
}
\vspace{0.5cm}
 \end{@twocolumnfalse}
  ]

\section{Introduction}

\footnotetext{\textit{$^{a}$~Theoretical Division, Los Alamos National Laboratory, Los Alamos, New Mexico 87545, USA. Fax: 1 505 606 0917; Tel: 1 505 665 1134; E-mail: cjrx@lanl.gov}}
\footnotetext{\textit{$^{b}$~Department of Physics, University of Notre Dame, Notre Dame, Indiana 46556, USA. }}

There are a wide variety of systems that exhibit pattern formation
in the form of ordered stripes, which can often be attributed to
the existence of competing interactions 
\cite{1,2,3,4,5,6,7,8}.
Such stripe morphologies appear in soft
matter systems such as colloids \cite{3,4,8,10,11,12,13,14} and
lipid monolayers near critical points \cite{15}; in
magnetic systems \cite{16}; in
certain superconducting vortex systems
such as low-$\kappa$ materials \cite{17},
multi-layered \cite{18,19} superconductors,
or multi-band superconductors \cite{20}; in charge-ordered states observed
in quantum Hall systems \cite{21} or cuprate superconductors \cite{22};
and in ordered states of dense
nuclear matter \cite{23}. 
Numerous studies have been devoted to understanding what types of
particle-particle interactions
can give rise to such patterns \cite{3,4,5,6,24,25,26,27}. Having a clear
methodology to control the patterns would be very useful for self-assembly and
tailoring specified morphologies for applications.

One aspect of these stripe-forming systems that has received
little attention is the effect 
on the pattern formation of adding a periodic substrate.
There are many examples
of systems in which the addition of a periodic substrate can
induce different types of ordering.  
The substrate may occur naturally at the atomic scale due to molecular
ordering at a surface,
or a substrate can be imposed using an external field or by 
nanostructuring or etching the surface.
A system of repulsively interacting colloids
forms a triangular lattice in the absence of a substrate, but when the 
colloids are placed on
an optically created quasi-one-dimensional (q1D) periodic substrate,
a number of distinct crystalline and smectic orderings appear as a function
of substrate strength or commensurability
\cite{28,29,30,31,32,33,34,35,36,37}.
Similarly,
magnetic colloids interacting with a fabricated q1D
corrugated surface \cite{38,39} also exhibit crystalline disordered and
smectic phases as a function of particle density \cite{38}.
In a superconducting vortex system, when
a q1D modulated substrate 
is created by etching
the surface of the superconductor, different types
of commensurability effects appear that 
are correlated with ordered and disordered vortex structures \cite{40,41,42}.

In this work we examine the two-dimensional ordering of particles with
long range repulsion and short range attraction interacting with a periodic
q1D substrate. The particular model we examine combines Coulomb
repulsion with a short-range exponential attraction between particles.
In the absence of a substrate, this system is known to exhibit
bubble, stripe, void, and uniform phases which have been well characterized
as a function of particle density and the ratio of attraction to repulsion
\cite{6,9,11,17,24,43}.
We specifically focus on parameter regimes in which the system
forms stripes in the absence of a substrate \cite{6}.
It might be expected that the addition of 
a q1D periodic substrate to a stripe system
would produce only a limited range of phases since the stripes could simply align
with the substrate; however,
we find that this
system exhibits a remarkably rich variety of distinct phases
as a function of substrate strength and the ratio of the 
particle spacing to the substrate minima spacing.
These phases include
2D modulated structures, prolate clump crystals, 
void crystals, and ordered kink arrays.
Additionally the stripes can be aligned perpendicular to the substrate
troughs.
Our results show that the addition of q1D substrates
can be a new route to controlling pattern formation in  systems with competing
interactions.

\begin{figure}
\centering
\includegraphics[width=0.5\textwidth]{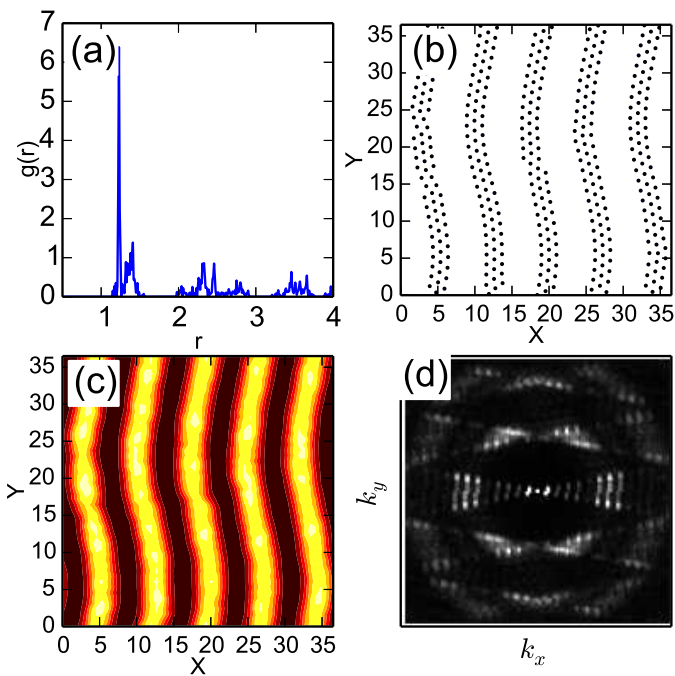}
\caption{
The stripe system in the absence of a substrate at a density of $\rho = 0.3$.
(a) Radial distribution function $g(r)$ showing a peak at $r = 1.2$.
(b) Real space image of the stripes. (c) Density plot in which
high densities correspond to brighter spots. (d) $S(k)$.
}
\label{fig:Fig1}
\end{figure}

\section{Simulation}   

We consider a two-dimensional system
with periodic boundary conditions of size
$L \times L$ containing
$N$ particles
that have pairwise interactions including
both repulsive and attractive components.
The particle configurations
are obtained by annealing the system from a high temperature
molten state in small increments to zero temperature. The particle
dynamics are governed by the following overdamped equation:
\begin{equation}
\eta \frac{d {\bf R}_{i}}{dt} =
-\sum^{N}_{j \neq i} \nabla V(R_{ij}) + {\bf F}^{s}_{i} +
{\bf F}^{T}_{i} .
\end{equation}
Here $\eta$ is the damping term which we set to unity and
${\bf R}_{i (j)}$ is the location of particle $i (j)$.
The particle-particle interaction potential
has the form $V(R_{ij}) = 1/R_{ij} - B\exp(-\kappa R_{ij})$,
where $R_{ij}=|{\bf R}_i-{\bf R}_j|$ and 
${\bf \hat R}_{ij}=({\bf R}_i-{\bf R}_j)/R_{ij}$.
The Coulomb term $1/R_{ij}$ produces a repulsive interaction at
long range, 
while the exponential term gives
an attraction at shorter range. 
At very short range
the repulsive Coulomb interaction becomes dominant again.
For computational efficiency, we employ a Lekner summation method to treat
the long-range Coulomb term \cite{longrange}.
The particle density is $\rho = N/L^2$, and unless otherwise noted
we take $\rho=0.3$.
In the absence of a substrate, previous studies of this model
found that for fixed $B = 2.0$ and $\kappa = 1.0$,
the system initially forms clumps at low density that grow in size
up to $\rho=0.27$.
For $0.27 < \rho \leq 0.46$ the system forms stripes, 
for $0.46 < \rho \leq 0.58$ void crystals form, 
and a uniform triangular lattice
appears for $\rho > 0.58$ \cite{6}.
Here we fix $B=2.0$ and $\kappa=1.0$
and focus on the stripe regime
near the density of $\rho = 0.3$ illustrated
in Fig.~\ref{fig:Fig1}(b), where an array of stripes forms
with three rows of particles in each stripe.
The stripe ordering is also apparent in the corresponding density plot of
Fig.~\ref{fig:Fig1}(c).
Figure~\ref{fig:Fig1}(a) shows the radial density function
$g(r)$ which has a first neighbor peak at $1.2a_{0}$, the value of
the intra-stripe particle distance $a_{\rm intra}$.
The structure factor $S(k)$ in
Fig.~\ref{fig:Fig1}(d) has six maxima regions
at large $k$ produced
by the tendency of the particles to form hexagonal structures
within each stripe.
The two bright peaks at small $k$ indicate the stripe ordering.
For the parameters we consider,
the interparticle potential has a minimum at $R_{ij} = 1.47a_{0}$. 
We focus on systems of size $L = 36.5a_{0}$.

\begin{figure}
\centering
\includegraphics[width=0.5\textwidth]{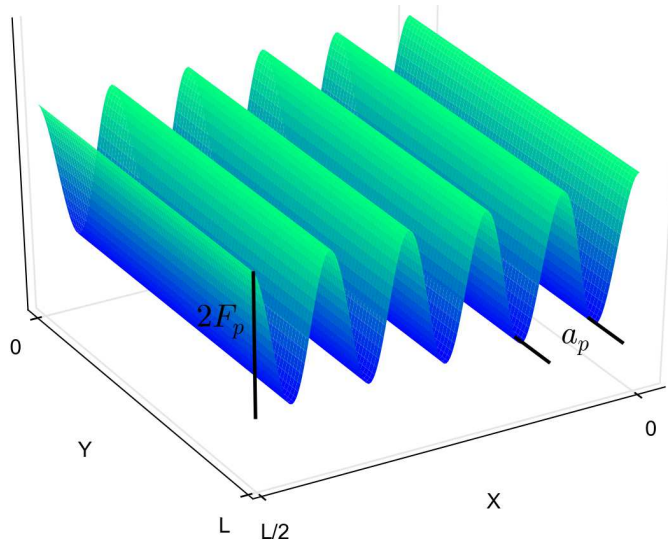}
\caption{
An example of a q1D periodic pinning substrate used in this work.  The 
potential is modulated in the $x$ direction and the substrate troughs are
parallel with the $y$ axis.  The total depth of each well is $2F_p$ and
the spacing between substrate minima is given by $a_p$.
}
\label{fig:Fig0}
\end{figure}

The force from the q1D pinning periodic substrate  ${\bf F}_{s}$ is
given by
\begin{align}
{\bf F}_s = F_p cos(2\pi x/a_p) 
\end{align}
where $a_{p} = L/N_p$, $N_{p} $ is the number of substrate minima, and
$a_{p}$ is the spacing between minima.
Such a substrate is illustrated in Fig.~\ref{fig:Fig0}.
The pinning force amplitude is $F_p$ and we consider values
in the range 
$0.01 \leq F_p \leq 6.0$.
Our primary interest is in the regime $F_p < 2.0$ since
the transition
from particle interaction-dominated to substrate
interaction-dominated behavior
typically occurs within this limit.
The thermal force ${\bf F}^T$ applied during the annealing phase
is modeled as Langevin kicks
with the properties
$\langle F^{T}(t)\rangle = 0$ 
and $\langle F^{T}_{i}(t)F^{T}_{j}(t^{\prime})\rangle = 2\eta k_{B}T
\delta_{ij}\delta(t - t^{\prime})$.  We start from a high temperature
liquid state and decrease the temperature in small increments until
$T = 0$, as in previous studies \cite{6}.

\begin{figure}
\centering
\includegraphics[width=0.5\textwidth]{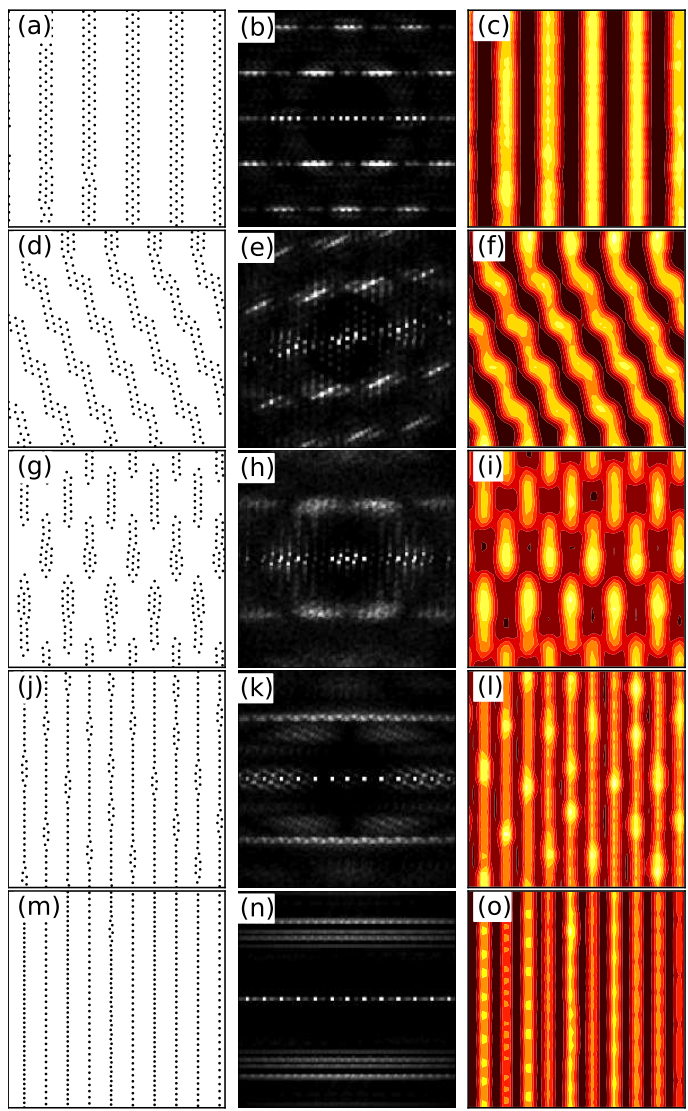}
\caption{
Real space particle positions (left column),
S(k) (central column), and density plots (right column)
for stripes ordering
on a periodic q1D substrate with troughs aligned in the $y$ direction
for a system with $a_{p} = 3.65$.
(a,b,c) Aligned stripe phase at $F_{p} = 0.05$. 
(d,e,f) Modulated stripes at $F_{p} =0.08$.
(g,h,i) Prolate clump phase at $F_{p} = 0.2$. 
(j,k,l) 1D kink phase at $F_{p} = 0.8$.
(m,n,o) Smectic phase at $F_{p} = 2.0$.
}
\label{fig:Fig2}
\end{figure}

\section{Results}

We first consider the case where the distance 
$a_p$ between the substrate minima
is significantly larger than the nearest-neighbor particle distance
$a_{\rm intra}$
within a stripe.
In Fig.~\ref{fig:Fig2} we show the particle positions, $S(k)$, and density plots
for a system with $a_{p} = 3.65a_{0}$, giving $a_{p}/a_{\rm intra} = 3.0$.
At $F_p=0.05$, 
Fig.~\ref{fig:Fig2}(a,b,c) indicates that the substrate aligns the stripes
along the $y$-direction, parallel to the substrate troughs. 
For $F_{p} < 0.06$ the
stripes remain aligned in the $y$ direction and half of the substrate
minima contain no particles,
since the substrate-free system
forms five stripes and there are ten substrate minima.
At $F_p=0.08$ in 
Fig.~\ref{fig:Fig2}(d,e,f), the stripes have tilted 
and develop an additional modulated structure in the form of steps.
These modulations have a tilted square
ordering which can be more clearly seen in the density plot of 
Fig.~\ref{fig:Fig2}(f).
At $F_{p} = 0.2$ in Fig.~\ref{fig:Fig2}(g,h,i), the stripes
break up and the system forms an  array of prolate clumps 
that have a 2D periodic ordering. This ordering produces
additional features in $S(k)$ at small $k$ values as shown in 
Fig.~\ref{fig:Fig2}(h).
The breaking apart of the original stripes permits each of the ten substrate
minima to capture an approximately equal number of particles.
The clumps exhibit some asymmetry, with the clump width varying from
three rows of particles at the center of Fig.~\ref{fig:Fig2}(g) 
to two rows of particles 
elsewhere. 
This produces a
smearing of the sixfold ordering at larger values of $k$ in 
Fig.~\ref{fig:Fig2}(h).
At $F_{p} = 0.8$ in Fig.~\ref{fig:Fig2}(j,k,l), stripe ordering returns when
the particles form nearly 1D chains stretching along the length of each
potential minima.
These 1D chains are interspersed with
kinks of smaller zig-zag patterns.  The kinks have an effective
repulsive interaction and tend to form a triangular lattice, as shown
in Fig.~\ref{fig:Fig2}(l).
As $F_{p}$ increases further,
the size and number of kinks
gradually deceases until the system forms a smectic state of
1D chains as shown in Fig.~\ref{fig:Fig2}(m,n,o) at $F_{p} = 2.0$. 
For further increases in $F_{p}$, we find
no changes in the smectic structure.

\begin{figure}
\centering
\includegraphics[width=0.5\textwidth]{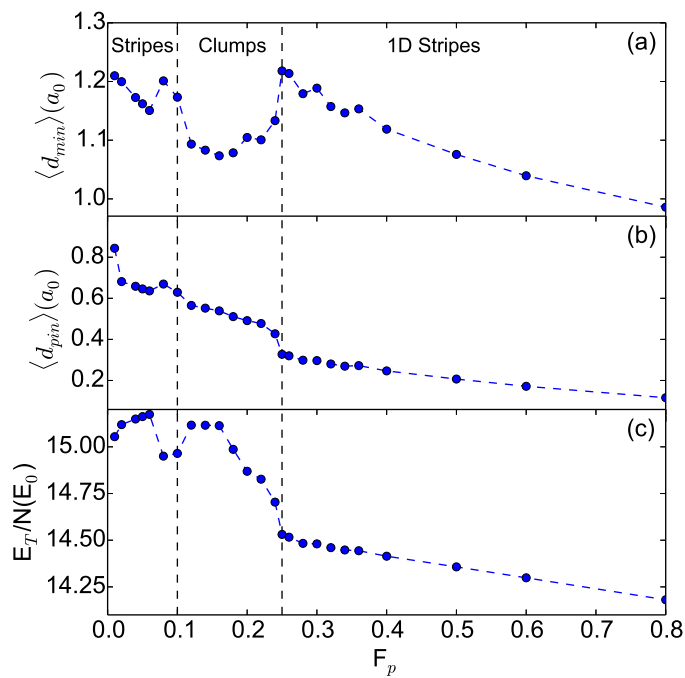}
\caption{
(a) Nearest neighbor distance $\langle d_{min}\rangle$, 
(b) horizontal distance to closet
substrate minima $\langle d_{pin}\rangle$, 
and (c) total energy of the system $E_{T}/N$ vs
$F_{p}$ for the system in Fig.~\ref{fig:Fig2}, 
highlighting the changes in the patterns.
}
\label{fig:Fig3}
\end{figure}

We characterize the onset of the different orderings by measuring the
average nearest-neighbor particle distance 
$\langle d_{min}\rangle = N^{-1}\sum^{N}_{i=0}d_{ni}$, where
$d_{ni}$ is the distance to the nearest neighbor of particle $i$ as
obtained from a Delaunay construction.
We also measure $\langle d_{pin}\rangle$ 
which is the average horizontal distance from a particle 
to the closest substrate
minimum, $\langle d_{pin}\rangle = N^{-1}\sum^{N}_{i=0}(x_i - x_p)$, 
where $x_i$ is the location in the
$x$ direction of particle $i$ and 
$x_p$ is the location of the nearest substrate minimum. 
If all the particles reside
at the substrate minima, $\langle d_{pin}\rangle = 0.$
We also measure  the total normalized energy of the system $E_{T}/N$.
In Fig.~\ref{fig:Fig3} we plot $\langle d_{min}\rangle$, 
$\langle d_{pin}\rangle$, 
and $E_{T}/N$ vs $F_{p}$ for the system
in Fig.~\ref{fig:Fig2}. 
There is a feature near $F_{p} = 0.08$ at the point where
the straight
stripes shown in Fig.~\ref{fig:Fig2}(a,b,c) 
transition to the modulated stripe phase shown in Fig.~\ref{fig:Fig2}(d,e,f). 
As $F_p$ further increases,
the modulated
stripes gradually transform into the clump phase 
shown in Fig.~\ref{fig:Fig2}(g,h,i). 
Near $F_p=0.28$ we find
a signature of the
transition from the clumps
to the 1D kinked stripe state shown in Fig.~\ref{fig:Fig2}(j,k,l) 
in the form of a peak
in 
$\langle d_{min}\rangle$, a dip in $\langle d_{pin}\rangle$, 
and a cusp in $E_{T}/N$.
For $F_{p} > 0.28$ the curves are smooth as
the number of kinks gradually decreases and the particles move closer 
to the substrate minima.  This is indicated by the steady decrease 
of $\langle d_{pin}\rangle$ which
approaches zero as the system forms the fully smectic state 
shown in Fig.~\ref{fig:Fig2}(m,n,o).

\begin{figure}
\centering
\includegraphics[width=0.5\textwidth]{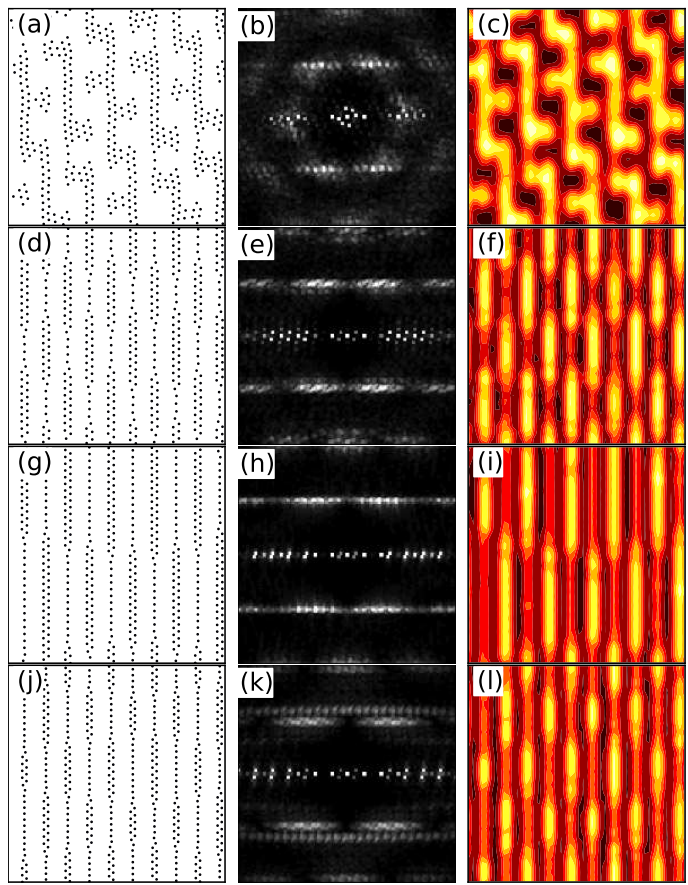}
\caption{
Real space particle positions (left column), $S(k)$ (central column), 
and density plots (right column) for a 
periodic q1D substrate with $a_{p} = 3.65a_{0}$ at $\rho = 0.363$.
(a,b,c) Modulated stripe phase at $F_{p} = 0.2$.
(d,e,f) Ordered kink phase at $F_{p} = 0.3$.
(g,h,i) Ordered kink phase at $F_{p} = 0.5$.
(j,k,l) Ordered kink phase at $F_{p} = 1.0$.
}
\label{fig:Fig4}
\end{figure}

For the same set of parameters but larger $a_p$ 
we observe the same
set of patterns. 
If we increase the particle density $\rho$ but hold the substrate period 
fixed, new patterns appear. 
At higher $\rho$ the 
prolate clump phase is lost but new types of modulated kink phases occur.
Fig.~\ref{fig:Fig4} shows the real space, $S(k)$, and density plots
for a system with $a_{p} = 3.65a_{0}$ at a particle density of 
$\rho=0.363$ where
the substrate-free system still forms stripes.  
The increase in particle density makes it more difficult to compress the
particles into the 1D patterns observed at $\rho=0.3$ in Fig.~\ref{fig:Fig2}.
Figure~\ref{fig:Fig4}(a,b,c) 
shows the ordering for $F_{p} = 0.2$, where a modified
stripe phase containing a semiperiodic array of spokes appears.  
At $F_p=0.3$, Fig.~\ref{fig:Fig4}(d,e,f) shows that
an ordered array of kinks forms where each substrate minimum contains
regions of two rows of particles interspersed
with regions that are only a single row wide. 
The density plot in Fig.~\ref{fig:Fig4}(f) indicates that the kinks order
into a periodic structure. 
As $F_{p}$ increases the number of kinks changes,
as shown in Fig.~\ref{fig:Fig4}(g,h,i) for $F_{p} = 0.5$. 
As $F_{p}$ is further increased
the system gradually develops more 1D behavior 
as illustrated in Fig.~\ref{fig:Fig4}(j,k,l) at $F_{p} = 1.0$.
For high enough $F_{p}$, all of the kinks vanish.

\begin{figure}
\centering
\includegraphics[width=0.5\textwidth]{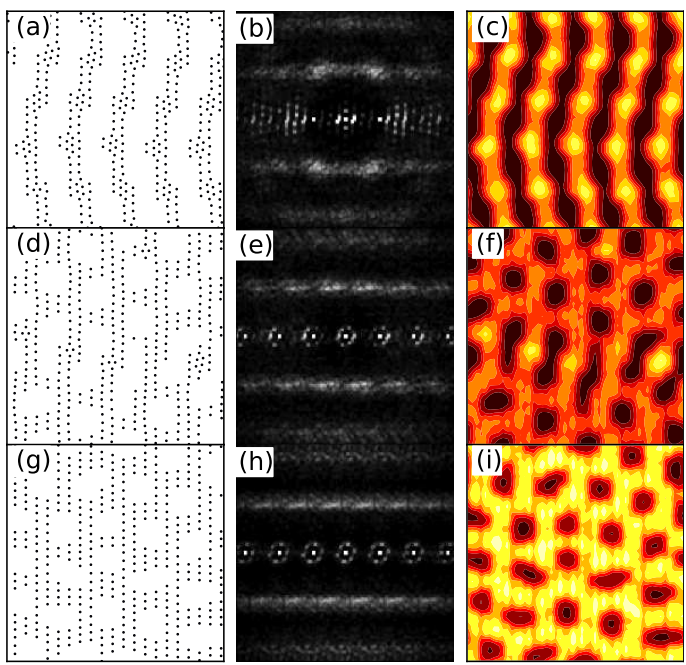}
\caption{
Real space particle positions (left column), $S(k)$ (central column), 
and density plots (right column) for
a periodic q1D substrate with $a_{p}  = 1.82$.
(a,b,c) Modulated stripe phase with a 2D periodic array of bubbles at
$F_{p}  = 0.14$.
(d,e,f) Void phase at $F_{p} = 0.2$.
(g,h,f) A better-defined void phase at $F_{p} = 2.0$.
}
\label{fig:Fig5}
\end{figure}

\begin{figure}
\centering
\includegraphics[width=0.5\textwidth]{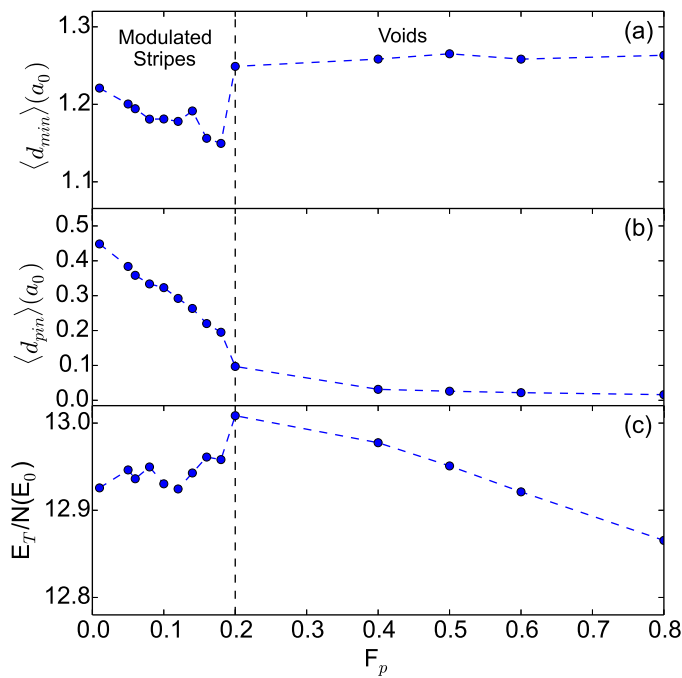}
\caption{
(a) $\langle d_{min}\rangle$, 
(b) $\langle d_{pin}\rangle$, and $E_{T}/N$ vs $F_{p}$ for the system in 
Fig.~\ref{fig:Fig5} showing the onset of the different phases.  
S: stripe phase; MS: modulated stripe phase.
}
\label{fig:Fig6}
\end{figure}

We next consider the limit in which the spacing 
$a_p$ between substrate minima becomes
comparable to or smaller than the average nearest-neighbor particle spacing 
$a_{\rm intra}$.  In Fig.~\ref{fig:Fig5} we plot
the real space particle positions, $S(k)$, and the local density for
samples with
$a_{p} = 1.82$ and $a_{p}/a_{intra} = 1.5$ for varied $F_{p}$. 
We find that when the pinning density is high, the original stripe
structure remains intact up to relatively large values of $F_p$ since
the smaller substrate spacing permits all of the particles to take
advantage of substrate minimum locations while still remaining in
the original stripe pattern.
Above this point,
as $F_{p}$ is increased we observe a modulated stripe phase with square ordering
as shown in Fig.~\ref{fig:Fig5}(a,b,c) for $F_{p} = 0.14$. 
At higher $F_p$ there is
a transition to a void crystal of the type shown in Fig.~\ref{fig:Fig5}(d,e,f) 
for $F_p=0.2$.
This void crystal becomes more stable 
and persists as $F_p$ is further increased,
as illustrated in Fig.~\ref{fig:Fig5}(g,h,i) for
$F_{p} = 2.0$.
In Fig.~\ref{fig:Fig6}(a,b,c) we 
plot the corresponding values of $\langle d_{min}\rangle$, 
$\langle d_{pin}\rangle$, and $E_{T}/N$  versus $F_{p}$ for the system
in Fig.~\ref{fig:Fig5}.  
At $F_{p} = 0.1$, there is an inflection in $\langle d_{min}\rangle$ 
at the transition from the
stripe to the modulated stripe phase.
The onset of the void phase 
near $F_p=0.2$ is marked by features in
$\langle d_{min}\rangle$.
Once the voids have formed, they remain stable for increasing $F_{p}$
since all the particles can fit in a potential minimum. 
We observe similar
void formation when we fix $a_p=1.82$ but vary the particle density $\rho$.

\begin{figure}
\centering
\includegraphics[width=0.5\textwidth]{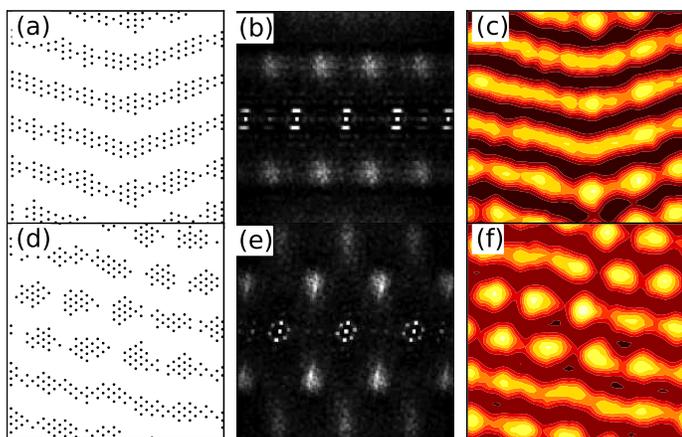}
\caption{
Real space particle positions (left column), $S(k)$ 
(central column), and density plots (right column) for periodic q1D substrates.
(a,b,c) A perpendicular stripe at $a_{p} = 1.2$ and $F_{p} = 2.0$. 
(d,e,f) A clump phase at $a_{p} = 0.9125$ and $F_{p} = 0.8$.
}
\label{fig:Fig7}
\end{figure}

When the pinning density is increased, the stripe state persists to
higher values of $F_p$.
The stripe alignment, however, alters and we find that the stripes
generally run perpendicular to the direction of the substrate troughs when
$a_p$ is small, as shown in
in Fig.~\ref{fig:Fig7}(a,b,c) for $F_{p}= 2.0$,
$a_{p} =  1.2$, and $a_{p}/a_{\rm inter} = 1.0$.
When $a_{p} < a_{\rm inter}$ we observe a transition from the stripe phase
to a clump phase as illustrated
in Fig.~\ref{fig:Fig7}(d,e,f) 
at $F_{p} = 0.8$ for a system with $a_{p}/a_{\rm inter} = 0.76$.
In general, for $a_{p}/a_{inter} < 1.0$, clump phases form at large $F_{p}$.

\section{SUMMARY}

We examine a stripe forming system
interacting with a periodic quasi-one dimensional
substrate.
We show that as a function of substrate strength and density,
a remarkably rich variety of distinct orderings can be realized.
These phases include stripes containing modulations that themselves form
a 2D ordered structure,
prolate clump phases, various types of 2D ordered
kink arrays, and smectic structures.
For denser substrate arrays we observe
transitions from a modulated stripe phase to a void crystal
or a clump phase.
Our results show that corrugated substrates could provide
a possible new route to controlling pattern forming systems.   

\section{Acknowledgements}
This work was carried out under the auspices of the 
NNSA of the 
U.S. DoE
at 
LANL
under Contract No.
DE-AC52-06NA25396.



\footnotesize{
 
}
\end{document}